\documentclass[12pt,reqno]{article}
\usepackage{graphicx}
\usepackage{amsmath}
\usepackage{times}
\usepackage{hyperref}
\usepackage{float}
\usepackage{accsupp}
\usepackage{fullpage}

\begin{document}

\title{
Dimensionality Reduction in Sentence Transformer Vector Databases with Fast Fourier Transform
\\
\small NextAI Systems \normalsize
}
\author{
Vitaly Bulgakov \href{mailto:vbulgakov@nextaisystems.com}
{\BeginAccSupp{method=escape,ActualText={}}vbulgakov@nextaisystems.com\EndAccSupp{}}
\and 
Alec Segal \href{mailto:alecsegal@gmail.com}
{\BeginAccSupp{method=escape,ActualText={}}alecsegal@gmail.com\EndAccSupp{}}
}

\date{\today}

\maketitle

\begin{abstract}
Dimensionality reduction in vector databases is pivotal for streamlining AI data management, enabling efficient storage, faster computation, and improved model performance. This paper explores the benefits of reducing vector database dimensions, with a focus on computational efficiency and overcoming the curse of dimensionality. We introduce a novel application of Fast Fourier Transform (FFT) to dimensionality reduction, a method previously underexploited in this context. By demonstrating its utility across various AI domains, including Retrieval-Augmented Generation (RAG) models and image processing, this FFT-based approach promises to improve data retrieval processes and enhance the efficiency and scalability of AI solutions. The incorporation of FFT may not only optimize operations in real-time processing and recommendation systems but also extend to advanced image processing techniques, where dimensionality reduction can significantly improve performance and analysis efficiency. This paper advocates for the broader adoption of FFT in vector database management, marking a significant stride towards addressing the challenges of data volume and complexity in AI research and applications. Unlike many existing approaches, we directly handle the embedding vectors produced by the model after processing a test input.  
\end{abstract}

\subsection*{1. Introduction}

Numerous studies are dedicated to enhancing computational efficiency in the context of deep learning models that generate embeddings. While we reference just a few most related to our study, it's important to note that a vast body of related research exists on this topic. \cite{PCA1} and \cite{PCA2} explore the use of Principal Component Analysis (PCA) on sentence embeddings to investigate their linguistic and semantic properties. 
Another recent study proposes a method called “Matryoshka Representation Learning”. MRL \cite{MRL} encodes information across multiple levels of granularity, enabling a singular embedding to adjust according to the computational demands of subsequent tasks. MRL seamlessly integrates into current representation learning workflows with negligible impact on inference and deployment costs. MRL learns coarse-to-fine representations that match or surpass the accuracy and depth of separately trained, lower-dimensional representations. \cite{MRL} also contains a rich bibliography of related studies.
\\\\
Unlike many existing approaches, we directly handle the embedding vectors produced by the model after processing a test input. Dense embedding models typically produce embeddings with a predefined number of dimensions, such as 768 or 1024. These vectors are then utilized in subsequent procedures, such as clustering, classification, semantic search, retrieval, and re-ranking. 
Estimating computational efforts for handling N-dimensional embedding vectors, such as those used in Retrieval-Augmented Generation (RAG) models, involves several factors including the dimensionality of the embeddings (\( \mathbf{N}\)), the number of embeddings or data points (\(\mathbf{M}\)), and the specific operations being performed (e.g., similarity search, clustering). The computational complexity can vary significantly based on the algorithm used for these operations. Here's a simplified estimation for a few common operations, such as Similarity Search with \(\mathbf{O(NM)}\) operations, Clustering with \(\mathbf{O(IKMN)}\), where \(\mathbf{I}\) is the number of iterations, K is the number of clusters, and sorting, like “mergesort” or “quicksort” with \(\mathbf{O(NMlog(M))}\) operations. As we can see, reduction of N can substantially reduce computational efforts.

\subsection*{2. Fourier Transform}

The Fourier Transform (FT) is a mathematical technique used to transform a signal from its original domain (often time or space) into a representation in the frequency domain. The fundamental idea behind the Fourier Transform is that any function (signal) can be represented as a sum of sinusoids (sines and cosines) of different frequencies, amplitudes, and phases. This transformation provides insights into the frequency components of the signal, which might not be apparent in its original form.

\subsubsection*{2.1 Continuous Fourier Transform}

The classical Continuous Fourier Transform of a continuous, time-dependent function $f(t)$ is defined as:

\[ F(\omega) = \int_{-\infty}^{+\infty} f(t) e^{-i \omega t} \,dt \]

where:
\begin{itemize}
    \item $F(\omega)$ is the Fourier Transform of $f(t)$,
    \item $t$ represents time,
    \item $\omega$ is the angular frequency (in radians per second),
    \item $i$ is the imaginary unit ($i^2 = -1$).
\end{itemize}

\subsubsection*{2.2 Inverse Fourier Transform}

The Inverse Fourier Transform allows recovering the original time-domain signal from its frequency-domain representation:

\[ f(t) = \frac{1}{2\pi} \int_{-\infty}^{+\infty} F(\omega) e^{i \omega t} \,d\omega \]
In our study we do not apply Inverse Fourier Transform, that will be explained further.

\subsubsection*{2.3 Discrete Fourier Transform (DFT)}

Discrete Fourier Transform, which is a discrete version of the FT, has a lot of applications where functions are represented as vectors of the Euclidean space. One of them is digital signal processing for signals sampled at discrete intervals. The DFT of a sequence of $N$ numbers $x_n$ is given by:

\[ X_k = \sum_{n=0}^{N-1} x_n e^{-i 2\pi k n / N} \]

where:
\begin{itemize}
    \item $X_k$ is the DFT output,
    \item $k$ is the index of the DFT output corresponding to the frequency component,
    \item $N$ is the total number of samples.
\end{itemize}

\subsubsection*{2.4 Fast Fourier Transform (FFT)}

The Fast Fourier Transform is an algorithm to compute the DFT efficiently, reducing the computational complexity from $O(N^2)$ for the direct evaluation of the DFT formula to $O(N \log N)$, making it significantly faster for large $N$. We will use FFT, which is implemented in various open source libraries including Python program language.

\subsection*{3. Method description}

The proposed algorithm was motivated by the following three observations:  

\begin{enumerate}
    \item The significance of low-frequency components in the Fourier Transform (FT) is particularly evident in cases where the underlying signal or data set exhibits slow-changing trends or patterns over time or space. These low-frequency components capture the fundamental, overarching structures within the data. This property is extensively used in various applications and analyses, such as Signal Processing and Filtering, Image Processing and Computer Vision, Data Compression, Economic and Financial Data Analysis, Climate and Environmental Studies, Biomedical Engineering, etc. In these cases, low-frequency components of the Fourier Transform are significant because they provide a clear representation of the main features and trends within the data.
    
    \item Solving problems in a transformed space, particularly using the Fourier Transform (FT) method, instead of directly tackling them in the original domain, offers significant advantages in various fields of science and engineering. The FT, by converting signals or data from the time or spatial domain into the frequency domain, unveils insights and properties that are not readily apparent or are more complex to deal with in the original domain. Examples where this approach is utilized are Simplification of Convolution Operations, Filter Design and Implementation, Spectral Analysis, Solution of Differential Equations and many others.  

    \item The training set and the language model architecture tend to accumulate substantial noisy information, which is manifested in the values of the embeddings vectors. Consequently, these vectors require some form of filtering. Intuitively, it is evident that this filtering or smoothing process can be effectively accomplished by eliminating the high-frequency components in the Fourier Transform. 
\end{enumerate}

In this article the object of our study is Large Language Models (LLM) and specifically Sentence Transformers \cite{ST}. Sentence Transformers are a type of model designed to convert sentences into high-dimensional vector representations, making them highly useful for a variety of natural language processing (NLP) tasks. These embeddings, essentially dense vectors, capture the semantic meaning of sentences in a way that allows for efficient computation of sentence similarities, information retrieval, and clustering, among other applications. Essential part here is played by BERT (Bidirectional Encoder Representations from Transformers) \cite{BERT}. BERT is a deep learning model designed for natural language processing. It works by pre-training on a large corpus of text, learning contextual relations between words in a sentence. In sentence transformers, BERT and its variations developed later, such as RoBERTa\cite{RoBERTa}, DistilBERT\cite{DistilBERT}, ALBERT\cite{ALBERT}, ColBERT\cite{ColBERT}, is adapted to produce sentence embeddings by processing entire sentences, leveraging its pre-trained contextual insights. The model captures the essence of each sentence in high-dimensional vector space, enabling various NLP tasks like semantic similarity, classification, and more. BERT's bidirectional nature allows it to understand the context of each word based on its surroundings, significantly enhancing the quality of the generated embeddings.  

Figure 1 illustrates, at a very high level, how two sentences are processed through the transformer and pooling layers, resulting in the conversion into vectors u and v. These vectors can then be tested for similarity.

\begin{figure}[H]
  \includegraphics[width=0.8\textwidth]{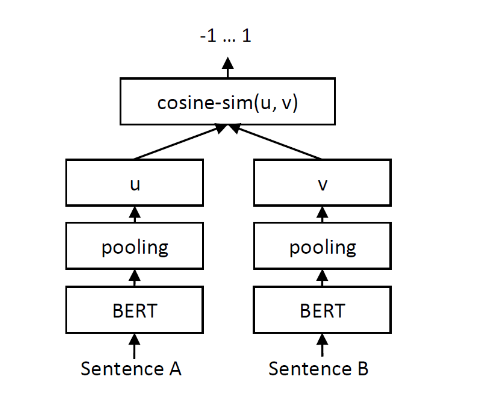}
  \caption{Two sentences are processed through the transformer and pooling layers}
  \label{fig:1}
\end{figure}

Let \( \mathbf{u} \) be the original embeddings vector of size \( N \), so \( \mathbf{u} \in \mathbf{R}^N \). Applying the Fourier Transform (FT) to \( \mathbf{u} \), we obtain the corresponding complex vector \( \mathbf{c} \) in the FT domain, where \( \mathbf{c} \) is of the same size \( N \). We define vector \( \mathbf{a} \) as a vector of FT amplitudes, which can be derived from Fourier coefficients:

\[ \mathbf{c} = \text{FT}(\mathbf{u}) \]

The amplitude of each component in \( \mathbf{a} \) is given by:

\[ A_k = \sqrt{\text{Re}(c_k)^2 + \text{Im}(c_k)^2} \]

for \( k = 0, 1, \ldots, N-1 \), where \( A_k \) is the amplitude of the \(k\)-th frequency component, and \( \text{Re}(a_k) \) and \( \text{Im}(a_k) \) are the real and imaginary parts of the \(k\)-th component, respectively.
\\\\
To perform dimensionality reduction by discarding higher frequency components, we create a new vector \( \mathbf{\hat{a}} \) of size \( M \) where \( M < N \) or even \( M \ll N \). This process can be represented as:

\[ \mathbf{\hat{a}} = \text{Reduce}(\mathbf{a}, M) \]

where \( \text{Reduce}(\mathbf{a}, M) \) denotes the operation of selecting the \( M \) most significant frequency components from \( \mathbf{a} \), effectively reducing its dimensionality from \( N \) to \( M \), and staying within the domain of amplitudes with a lower dimension \( M \). 

This approach yields two vector spaces: the original of size \( N \) and the one based on the FT domain of size \( M \). Each forms its own vector database, \(DB{orig}\) and \(DB{trans}\), where the latter is \(N/M\) times smaller. Both databases are created from the same set of documents, which will be illustrated further. We then compare retrieval results obtained by searching through these databases for the same query, converted to the embeddings vector and FT-transformed for the search in \(DB{trans}\) database.

\subsection*{4. Implementation}

The source of Sentence Transformers for our study was Hugging Face hub. We evaluated quite a few “sentence similarity” models and found that “all-mpnet-base-v2” one works best for our purposes. Its sequence size is 384 and embeddings dimension is 768.
The following code snippet illustrates how to get embeddings vectors from given sentences or documents:

\begin{verbatim}
from sentence_transformers import SentenceTransformer
sentences = [doc_1, doc_2, …, doc_n]
model = SentenceTransformer(model_name)
embeddings = model.encode(sentences)
\end{verbatim}

We convert embeddings vectors using FFT "fft" module from scipy.fft. The following code snippet illustrates how to convert a given embeddings vector, take amplitudes as an FF image space and retain M first (low frequency) components from the resulting image:
\begin{verbatim}
import numpy as np
from scipy.fft import fft
fft_result = fft(embeddings_vector)
images = np.abs(fft_result)
images_m = images[:M]
\end{verbatim}

All documents converted to embeddings vectors by the sentence transformer model are stored in Facebook FAISS vector database \cite{FAISS}. We retrieve most relevant (semantically close) to a query documents represented by vectors from this database index.  

\subsection*{5. Numerical examples}

We will use python's sklearn MDS (Multidimensional Scaling) module to show distances between vectors projected into 2D space. This technique is employed to project high-dimensional data onto a lower-dimensional space while preserving the pairwise distances between the data points as much as possible.
To illustrate how this works we will first use the following document where the first five paragraphs (P1 to P5) are dedicated to "machine learning" followed by five (P6 to P10) dedicated to "wine tasting". The entire text snippets were randomly generated by a language model without manual interaction:\\
\begin{scriptsize}
\par P1: Machine learning, a subset of artificial intelligence, focuses on building systems that learn from data to make predictions or decisions without being explicitly programmed for specific tasks. It's revolutionizing industries by enabling smarter, more efficient operations.

P2: The core idea behind machine learning is the development of algorithms that can process large sets of data, identify patterns, and make predictions. Techniques like neural networks, decision trees, and support vector machines are commonly employed in various applications.

P3: One significant application of machine learning is in natural language processing (NLP), where machines understand, interpret, and generate human language. This technology powers virtual assistants, chatbots, and translation services, enhancing communication across languages.

P4: Machine learning also plays a crucial role in computer vision, allowing computers to interpret and make decisions based on visual data. This capability is fundamental for facial recognition systems, autonomous vehicles, and medical imaging analysis.

P5: Advancements in deep learning, a subset of machine learning, are driving breakthroughs in areas like generative models and reinforcement learning. These innovations pave the way for more sophisticated AI applications, shaping the future of technology.

P6: Wine tasting is an art form that involves the sensory evaluation of wine, focusing on its color, aroma, and flavor profile. Enthusiasts and professionals alike engage in this practice to appreciate wine's complexity.

P7: The color of wine can provide insights into its age, grape variety, and production methods. Observers evaluate wine by tilting the glass against a white background to examine its hue and clarity.

P8: Aroma is a critical component of wine tasting, revealing information about the wine's grape variety, terroir, and aging process. Tasters often swirl the wine to release its bouquet before inhaling deeply to identify specific scents.

P9: Tasting wine involves evaluating its palate, noting the balance between sweetness, acidity, tannins, and alcohol. This stage assesses the wine's flavor profile, structure, and finish, contributing to its overall quality assessment.

P10: Wine tastings often occur in vineyards, wine bars, and festivals, serving as educational experiences and social gatherings. Participants learn to distinguish between different wine varieties and regions, enhancing their appreciation for this sophisticated beverage.
\end{scriptsize}
\\\\
We take "Training data in Machine Learning" as a query that obviously belongs to "machine learning" topic.
We first calculate distances between the "query vector" and other vectors corresponding to the above 10 paragraphs. All vectors have been transformed with FFT and reduced in size by a factor of 5 rounded to the closest integer. So the size of these vectors has been reduced from 768 to 153. Figure 2 illustrates distances between the query vector and paragraph vectors.

\begin{figure}[H]
  \includegraphics[width=0.8\textwidth]{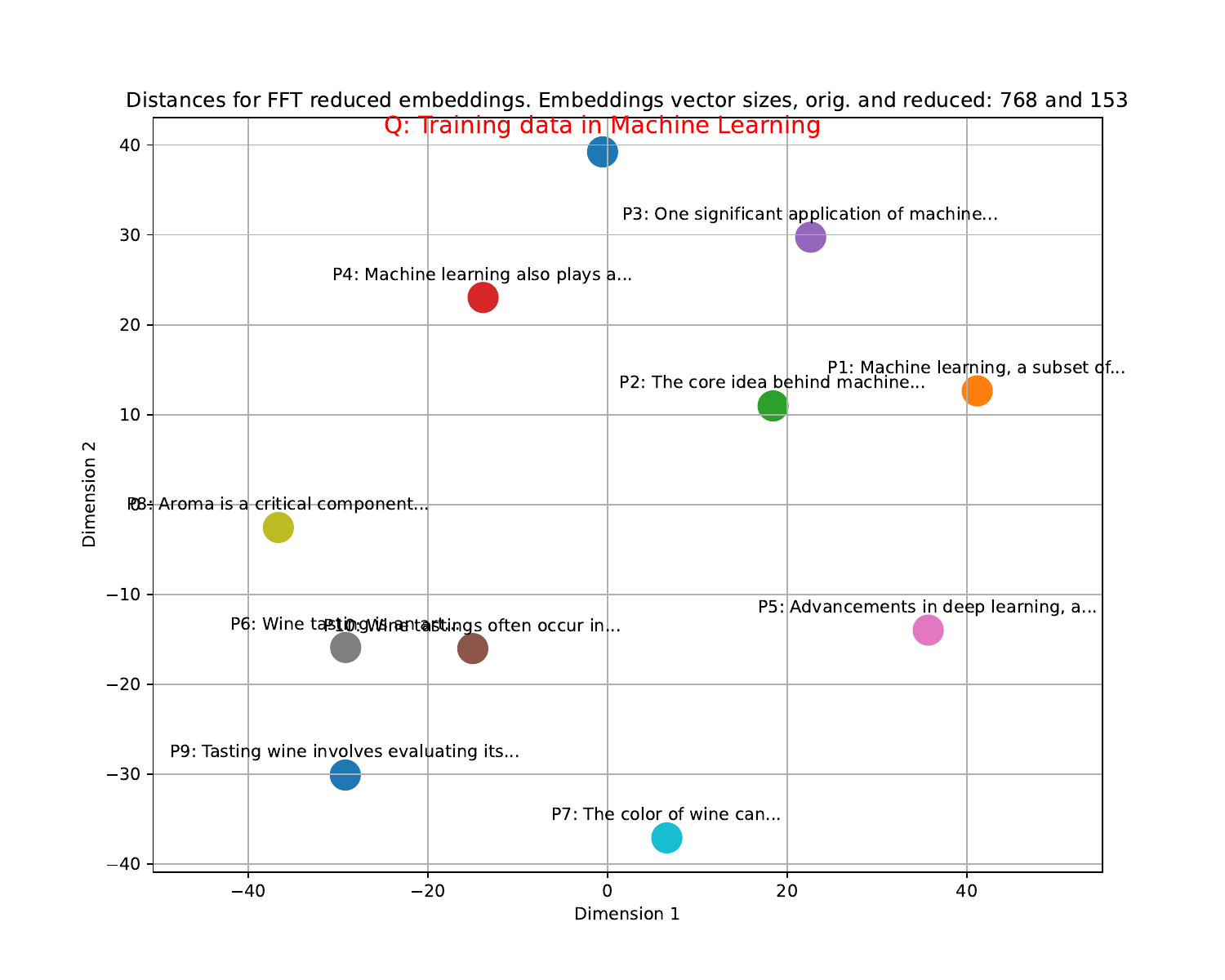}
  \caption{After retrieval vectors have been grouped into 2 clusters, "machine learning" and "wine tasting"}
  \label{fig:2}
\end{figure}

We can see that all 11 vectors (1 for a query and 10 for paragraphs) have been grouped in two cluster corresponding to "machine learning" and "wine tasting" topics respectively, as expected. If we try to retrieve only the first 4 documents, sorted by ascending order of distances, we will see that all of them belong to the right topic, see Figure 3.

\begin{figure}[H]
  \includegraphics[width=0.8\textwidth]{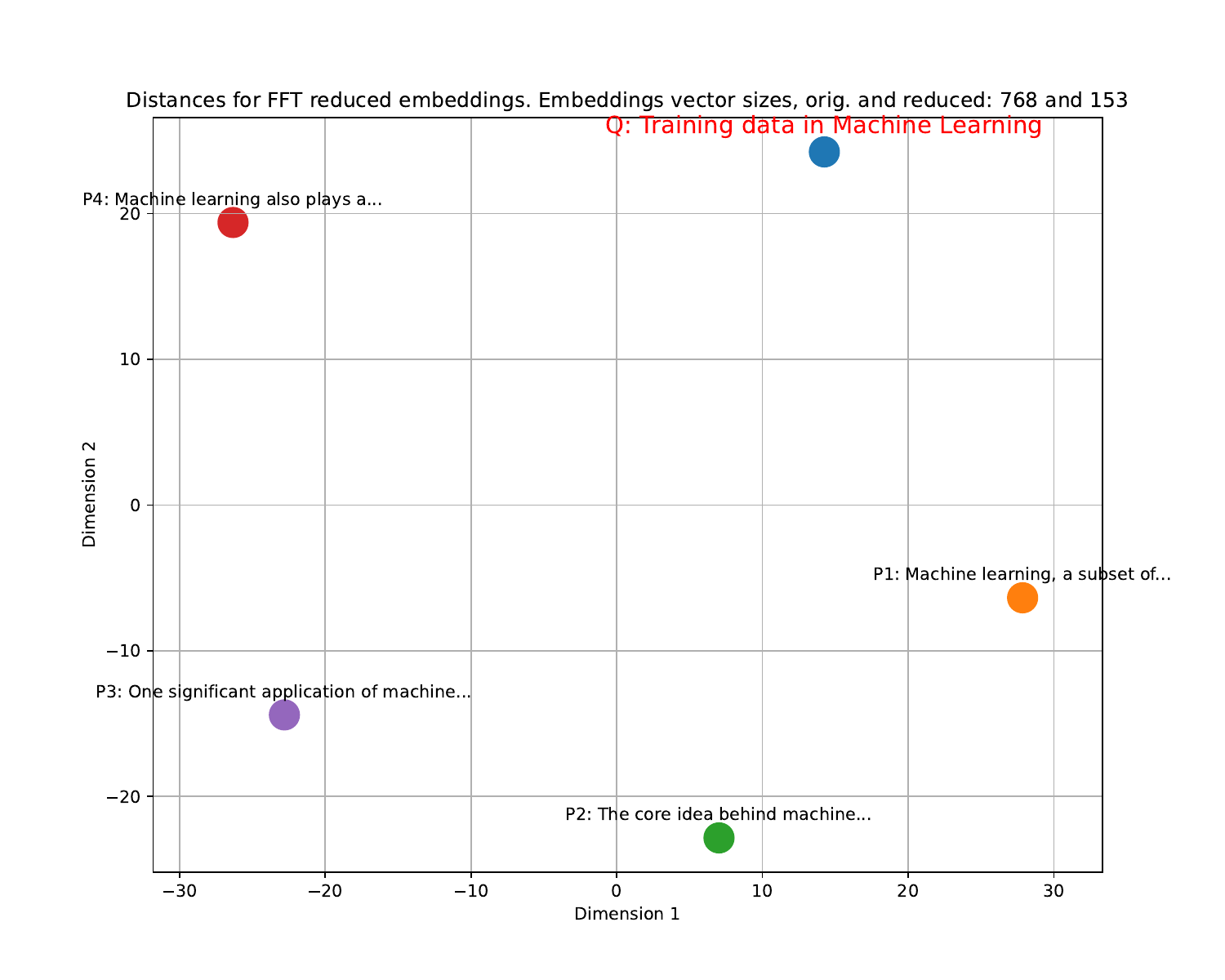}
  \caption{Reduction by order of 5 rounded to the closest integer. All first 4 retrieved vectors are related to "machine learning" topic}
  \label{fig:3}
\end{figure}

Let's further reduce transformed vectors dimension, this time by a factor of 8. This results in original and reduced vector sizes of 768 and 96, respectively. As we can see from Figure 4, all four retrieved vectors still belong to "machine learning" topic. After 10 times reduction we will get 1 irrelevant vector.

\begin{figure}[H]
  \includegraphics[width=0.8\textwidth]{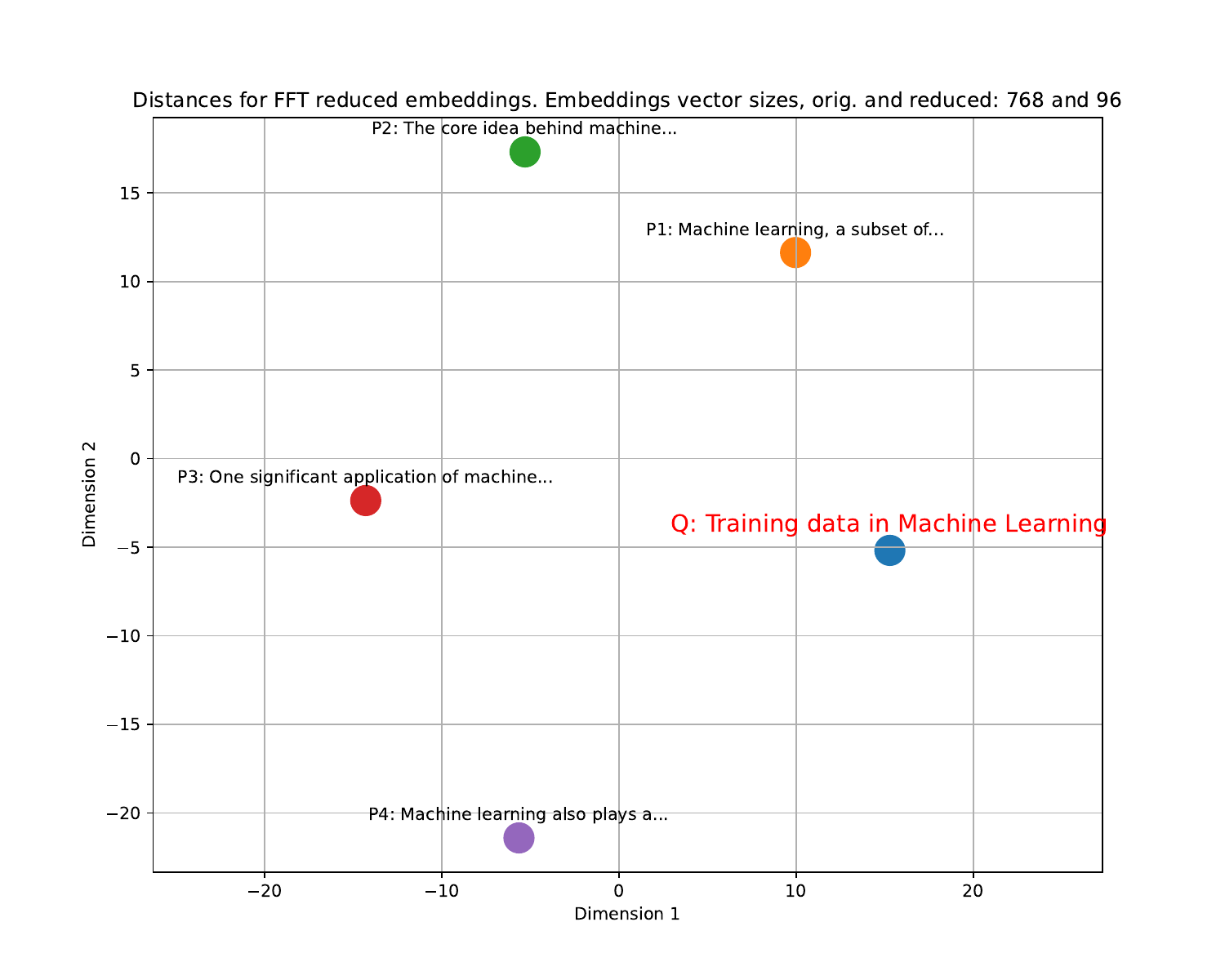}
  \caption{Reduction by order of 8. All first 4 retrieved vectors are related to "machine learning" topic}
  \label{fig:4}
\end{figure}

In the next example we examine a significantly larger and real-world dataset. It is a multi-volume publication entitled "The General Laws of Massachusetts" (\href{https://malegislature.gov/Laws/GeneralLaws/}{\it{General Laws of Massachusetts}}), that contains 602 chapters. The model's sequence length is set to 384. To determine an appropriate chunk size that should stay within the maximum sequence length, we divided 384 by 3 to split all content into chunks. We should 
mention here that our chunking strategy was to split the text by a  \textbackslash{n} splitter to prevent the random text segmentation. The flip side of this approach is that some of the textual information is lost as the chunk size exceeds the model sequence length.
Consequently, this led us to a total of 96,415 documents and 6,893,942 tokens processed by the model to create a vector database.\\
Let's see what documents will be retrieved on query "Q: Tell me about environmental protection in Massachusetts". 
We keep the last reduction factor of 8 and let's consider the first retrieved documents sorted by  ascending order of distances or descending order of similarity respectively. These distances are depicted in Figure 5. 

\begin{figure}[H]
  \includegraphics[width=0.8\textwidth]{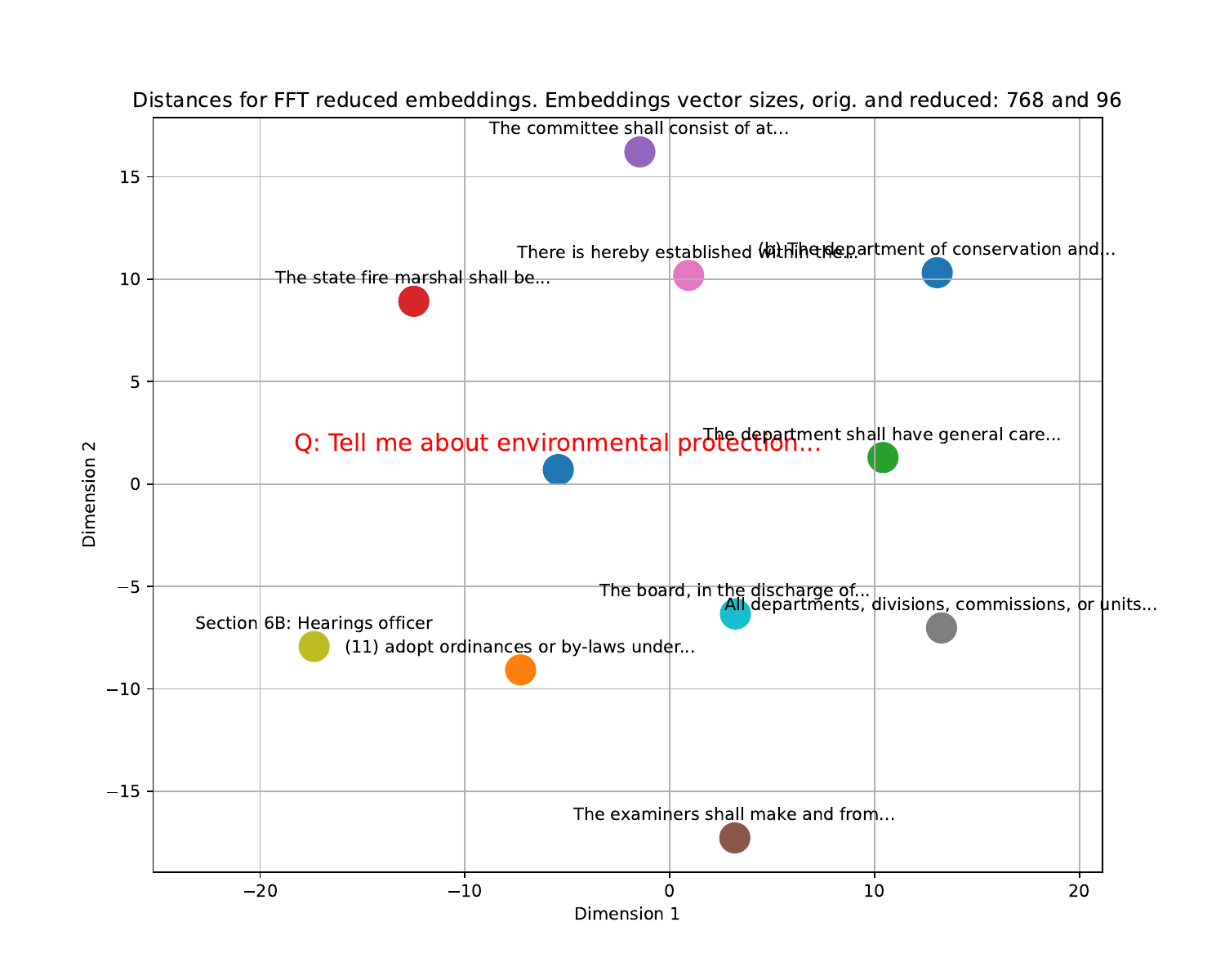}
  \caption{First documents retrieved from "The General Laws of Massachusetts" on "Q: Tell me about environmental protection in Massachusetts"}
  \label{fig:5}
\end{figure}

Since the diagram shows only very short snippets of retrieved documents and documents themselves could be relatively large, we will explain in few words what each of document is about:\\
\\ Doc 1 - is about regulating the design construction and use of buildings.
\\ Doc 2 - states that the department of environmental protection shall protect the interests of the commonwealth.
\\ Doc 3 - is about the Massachusetts fire training council and Underground Storage Tank Petroleum Product Cleanup.
\\ Doc 4 - is about financial interest in hazardous waste disposal.
\\ Doc 5 - is about examiners duties and regulations related to liquefied petroleum gas fitting containers, design standards to prevent fire, explosion, etc.
\\ Doc 6 - is about the executive office of environmental affairs and the board of registration of hazardous waste site cleanup professionals.
\\ Doc 7 - is about regulations for departments of ecology and marine resources of the waters of the sanctuary.
\\ Doc 8 - is about cooperative efforts in certifying the construction, operation and maintenance of energy facilities
\\ Doc 9 - stated that the department of conservation and recreation shall impose a surcharge of \$1 upon each fee charged and collected from admission to and parking in the Horseneck Beach Reservation
\\\\ As we can see all of them are about regulations and procedures related to state environment protection.

\section*{Conclusion}
In this study, we introduce a method that demonstrates strong performance across a wide array of both test and real-world datasets. However, accurately delineating the method's application boundaries proves to be a significant challenge. This issue is not unique to our approach but is common across the field of deep learning, where many effective methods and models are often developed based on intuitive reasoning. \\
An intriguing avenue for future exploration is to elucidate why the properties of the Fourier Transform affect embedding vectors similarly to their impact on time-dependent functions. We aim to continue this research to gain a deeper understanding of these phenomena.\\
One more property, that was observed from our numerical study, is that transformed vectors of reduced size can keep semantic meaning of the portion of text sometimes even better that original vectors. Our example with "The General Laws of Massachusetts" data showed that it highlights the resemblance between two segments of text that share a limited number of identical words, yet maintain closely related meanings.

\end{document}